\documentclass[aps,prl,twocolumn,showpacs,superscriptaddress]{revtex4}
\usepackage{rotating}
\usepackage{graphicx}
\begin{document}
\title{Network Harness: Metropolis Public Transport}
\author{C. von Ferber}
\email[]{ferber@physik.uni-freiburg.de}
 \affiliation{Complex Systems Research
Center, Jagiellonian University, 31007 Krak\'ow, Poland}
\affiliation{Physikalisches Institut, Universit\"at Freiburg,
79104 Freiburg, Germany}%
\author{T.~Holovatch}
 \affiliation{Ivan Franko National
University of Lviv, 79005 Lviv, Ukraine}
\author{Yu.~Holovatch}
  \affiliation{Institute for Condensed
Matter Physics of the National Academy of Sciences of Ukraine,
79011 Lviv, Ukraine} \affiliation{Institut f\"ur Theoretische
Physik, Johannes Kepler Universit\"at Linz, 4040 Linz, Austria}
\affiliation{Ivan Franko National University of Lviv, 79005 Lviv,
Ukraine}
\author{V.~Palchykov}
  \affiliation{Ivan Franko National
University of Lviv, 79005 Lviv, Ukraine}
\begin{abstract}
We analyze the public transport networks (PTNs) of a number of major
cities of the world. While the primary network topology is defined by
a set of routes each servicing an ordered series of given stations, a
number of different neighborhood relations may be defined both for the
routes and the stations.  The networks defined in this way display
distinguishing properties, the most striking being that often several
routes proceed in parallel for a sequence of stations. Other networks
with real-world links like cables or neurons embedded in two or three
dimensions often show the same feature - we use the car engineering
term {\em harness} for such networks. 
Geographical data for the routes reveal surprising self-avoiding walk
(SAW) properties. We propose and simulate an evolutionary model of PTNs 
based on effectively interacting SAWs that reproduces the key features.
\end{abstract}
\maketitle

\section{Introduction}\label{I}

Taken the general interest in networks of man-made and natural
systems \cite{reviews}, it is remarkable that one of the most
commonly encountered networks, the urban public transport network
(PTN), is much less studied. The PTN constitutes an
example of a transportation network and shares general features of
these systems: evolutionary growth, optimization, embedding into
two-dimensional (2D) space. However, as compared with other
transportation networks like airport networks \cite{Amaral00,air},
railway networks \cite{Sen03}, or power grid networks
\cite{Amaral00,powergrid} much less is known about the statistical
and topological properties of PTNs.

The few studies that have considered these questions so far have
either been performed on a subnetwork (tram, subway) of a specific
city like Boston or Vienna \cite{Marchiori00,Latora,Seaton04} or
they were limited by the number of cities or their size. The
recent thorough study of the bus and tram networks of 22 Polish
cities considered PTNs with up to 2811 stations
\cite{Sienkiewicz05}. Our preliminary work on some larger networks
\cite{Ferber05} was restricted to only three cities (Berlin,
D\"usseldorf, and Paris). The present picture that emerges from
this research views PTNs as networks with small-world properties
and hierarchical organization as derived from various network
characteristics and their correlations \cite{Sienkiewicz05}. While
indications for scale-free behavior have been found these depend
strongly on the interpretation of the network.

The present article is based on a survey of selected major cities of
the world with PTN sizes ranging between 2000 and 46000 stations.
Simulations of an evolutionary growth model based on self-avoiding
walks that we propose appear to reproduce most of the key features
of these PTNs. In addition to the standard characteristics of
complex networks like the number of nearest neighbors, mean path
length, and clustering we observe features specific to PTNs due to
their embedding in geographic space constrained by city structure.
The most striking being that often several routes proceed in
parallel on a common road or track for a sequence of stations.
While other networks with real-world links like cables or neurons
embedded in two or three dimensions often show similar behavior,
these can be studied in detail in our present case.

%
%
\begin{figure}[ht]
\includegraphics[width=8cm]{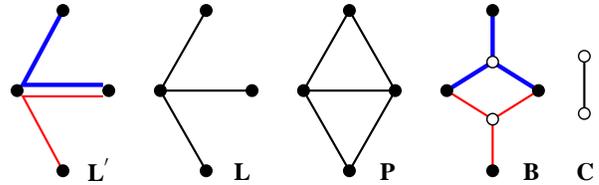}
   \caption{PTN graph representations.
   {\bf L$^{'}$}-space:
   filled circles represent stations serviced by two different
   routes shown by a bold and a thin line. {\bf L}-space:
   reduction of {\bf L$^{'}$} to a simple graph with single links.
   {\bf P}-space: any two stations that are serviced by a common
   route are linked. {\bf B}-space: stations  are  linked to the
   routes (open circles) they belong to. {\bf C}-space:
   routes are connected when they share a common station.
   \label{fig1}}
\end{figure}

From maps of PTNs it is obvious that routes in general do not follow 
the shortest path between their terminal stations but rather meander
through neighborhoods and between sites of interest. We quantify this 
behavior by analyzing geographical data. This leads us to model the
routes as self-avoiding walks that efficiently cover the surface 
while the sites of interest introduce an effective attraction between
them.

To work out the specific features of PTNs, it is natural to interpret the public
transport system of a city in terms of graphs, see Fig.
\ref{fig1}. The primary network representation, which we call {\bf
L$^{'}$}-space is defined by a set of routes each servicing an
ordered series of stations. For the PTNs studied so far, two
different neighborhood relations were used that lead to different
representations. In the first one, called {\bf L}-space, two
stations are defined as neighbors only if one station is the
successor of the other one in the series of stations serviced by a
route \cite{Latora}. In the second one, the {\bf P}-space, two
stations are neighbors whenever they are serviced by a common
route \cite{Sen03}. Some of the specific features found in PTNs
are more naturally described in additional representations that we
introduce. A bipartite graph, {\bf B}-space, is constructed by
representing both routes and stations as nodes of different type
as depicted by filled and open circles in Fig. \ref{fig1}. The
one-mode projection of the {\bf B}-space graph to the station
nodes leads back to the {\bf P}-space representation, whereas the
corresponding projection to the set of routes results in a
complementary {\bf C}-space graph. Note that {\bf L$^{'}$} differs
from {\bf L} only by the presence of multiple links. In a similar
fashion one may define additional primed spaces {\bf P$^{'}$},
{\bf C$^{'}$} by keeping multiple links also in {\bf P} and {\bf
C} \cite{Ferber06b}.

\begin{widetext}

 \begin{table}[htp]
\centering
\begin{tabular}{|l|r|r|c|r|c|c|c|c|r|c|c|r|c|r|r|r|r|c|}
\hline \hline City& $N$ & $M$ & $\kappa$ & $\kappa_{\rm p}$ &
$\gamma$ & $\hat{k}$ & $\hat{k}/\bar{k} $ & $\gamma_{\rm p}$ & $\hat{k}_{\rm p}$ &
$\hat{k}_{\rm p}/\bar{k}_{\rm p}$ &
$\overline{C}$ & ${\cal C}$ & $\overline{C_{\rm p}}$ & ${\cal C}_{\rm p}$& 
$\hat{\ell}$ & $\bar{\ell}$& $\hat{\ell}_{p}$ &
$\overline{\ell_{\rm p}}$\\
\hline
         Berlin &  2997 &  218 & 3.16 &  80.2 & (4.30) & 1.24 & 0.49 & (5.86) &  38.5 & 0.70 &  0.08 & 91.58 &  0.81 & 43.22 &  88 & 21.60 &   6 &  3.10 \\
         Dallas &  7163 &  131 & 2.22 & 136.3 & 4.99 & (1.01) & 0.49 & (4.67) &  76.9 & 0.77 &  0.01 & 37.87 &  0.97 & 63.34 & 269 & 85.80 &  10 &  3.78 \\
   D\"usseldorf &  1615 &  124 & 3.20 &  90.3 & (3.99) & 1.12 & 0.44 & (4.63) &  58.8 & 1.02 &  0.04 & 22.91 &  0.79 & 20.99 &  56 & 13.18 &   5 &  2.58 \\
        Hamburg &  8159 &  708 & 3.25 &  78.4 & (4.70) & 1.47 & 0.56 & (4.92) &  55.6 & 1.11 &  0.08 & 255.59 &  0.82 & 133.99 & 158 & 39.71 &  11 &  4.78 \\
      Hong Kong &  2117 &  321 & 5.22 & 230.1 & (3.04) & 2.60 & 0.72 & (4.40) & 125.0 & 1.01 &  0.16 & 92.33 &  0.73 & 12.51 &  60 & 11.11 &   4 &  2.26 \\
       Istanbul &  4043 &  414 & 2.69 & 140.1 & 4.04 & (1.13) & 0.49 & (2.70) &  71.4 & 0.93 &  0.03 & 44.99 &  0.79 & 41.54 & 131 & 29.69 &   6 &  3.09 \\
         London & 11026 & 2005 & 3.23 & 213.9 & 4.58 & (1.46) & 0.48 & 4.39 & (142.9) & 1.27 &  0.16 & 658.15 &  0.70 & 78.51 &  77 & 22.03 &   6 &  3.19 \\
    Los Angeles & 46244 & 1893 & 2.73 & 154.4 & 4.88 & (1.50) & 0.64 & 3.92 & (200.0) & 2.07 &  0.03 & 599.68 &  0.90 & 430.68 & 247 & 43.55 &  14 &  4.60 \\
         Moscow &  3756 &  679 & 7.94 & 129.7 & (3.31) & 2.12 & 0.65 & (2.91) &  50.0 & 0.78 &  0.11 & 129.78 &  0.74 & 43.14 &  28 &  7.07 &   5 &  2.52 \\
          Paris &  4048 &  232 & 6.41 &  79.5 & 2.61 & (3.24) & 0.94 & 3.70 & (100.0) & 2.07 &  0.07 & 86.78 &  0.88 & 72.88 &  47 &  7.22 &   5 &  2.79 \\
           Rome &  6315 &  681 & 3.02 &  86.4 & 4.39 & (1.16) & 0.45 & (5.87) &  45.5 & 0.76 &  0.03 & 69.19 &  0.73 & 76.93 &  93 & 29.64 &   8 &  3.58 \\
    S\~ao Paolo &  7223 &  998 & 5.95 & 333.6 & 2.72 & (4.20) & 1.29 & (3.06) & 200.0 & 1.46 &  0.23 & 514.11 &  0.73 & 38.32 &  33 & 10.34 &   5 &  2.66 \\
         Sydney &  2034 &  596 & 4.35 &  73.2 & 3.99 & (1.82) & 0.55 & (5.66) &  38.5 & 0.91 &  0.14 & 87.66 &  0.73 & 34.92 &  35 & 12.71 &   7 &  3.03 \\
         Taipei &  5311 &  389 & 4.02 & 415.5 & (3.74) & 1.75 & 0.56 & (5.17) & 200.0 & 0.85 &  0.11 & 188.89 &  0.69 & 15.38 &  74 & 20.86 &   6 &  2.35 \\
\hline
  $a$=0,$b$=0.1 &   635 &  500 & 2.77 & 216.6 & -- & -- & -- & 4.43 &  (76.9) & 0.47 &  -- &  -- &  0.76 &  2.9 & 110 & 36.10 &   4 &  1.96  \\
  $a$=0,$b$=0.5 &  3336 &  500 & 3.14 & 302.7 & -- & -- & -- & (7.66) & 111.1 & 0.62 &  -- &  -- &  0.68 & 12.6 & 190 & 49.21 &   7 &  3.00  \\
    $a$=0,$b$=8 &  5464 &  500 & 3.36 & 233.8 & -- & -- & -- & (12.35) &  71.4 & 0.46 & -- &  -- &  0.65 & 22.8 & 229 & 59.37 &   9 &  3.71  \\
\hline \hline
\end{tabular}

\caption{PTN characteristics  in {\bf L}- and {\bf
P}-representations. Index 'p' indicates {\bf P}-space
characteristics. The last three rows show data for simulated cities.
We list the number of stations $N$, the number of routes $M$, the
ratio $\kappa$ of the second $\overline{k^2}$ to the first moment
$\bar{k}$ of $p(k)$, the exponent $\gamma$, and the scale $\hat{k}$ of
fits of $p(k)$ to power and exponential laws and the ratio
$\hat{k}/\bar{k}$, the average clustering coefficient $\overline{
C}$ and its ratio ${\mathcal C}$ to the ER-value $C_{\rm
ER}=\bar{k}/N$, the maximal $\hat{\ell}$ and the mean $\overline{\ell}
$ shortest path lengths; e.g.. an average trip between the 11026 London stations needs $\overline{\ell_{\rm p}}$=3.2 changes with a maximum of $\hat{\ell}_{\rm p}$=6; see text. \label{tab1}}
\end{table}

\end{widetext}

For our empirical survey we acquired publicly available data of the
PTNs of 14 major cities from the web-pages of local public
transport organizations \cite{note1}. Table \ref{tab1} lists
cities together with some of the characteristics extracted from
our analysis. A more complete account will be given in a separate
publication \cite{Ferber06b}. To check for the small-world
properties of the networks we have analyzed the mean and maximal
shortest path lengths $\overline{\ell}$, $\hat{\ell}$.
The data for space {\bf L} given in Table \ref{tab1} show that
these numbers are very small as compared to the number of nodes.
 In the {\bf P}-space the shortest path
lengths $\overline{\ell_{\rm p}}$, $\hat{\ell}_{\rm p}$
correspond to the number of changes one should make traveling
between two given stations. From these data we conclude that  a
typical station is within relatively short reach from all other
stations. As has been shown \cite{Watts98}, small world networks
are also highly clustered as characterized by the clustering
coefficient. The latter is defined by $C_i = 2y_i/k_i(k_i-1)$
where $k_i$ is the degree of node $i$ and $y_i$ is the number of
links between its neighbors. The large ratios ${\mathcal C}=\overline{ C}/C_{\rm ER}$ of
the average values with respect to those of
Erd\"os-R\'enyi (ER) random graphs of the same sizes confirm the
high clustering that is present in these networks.

A very fruitful concept that has lead to a unifying view on
complex networks and also leads to their classification is that of
scale-free behavior \cite{reviews}. If $p(k)$, the probability
that given node of the network has degree $k$, follows a power law
$p(k)\sim k^{-\gamma}$, indicating a fat tail of the node degree
distribution, the network is said to be scale-free. While many
networks and in particular random graphs have an exponentially or
even faster decaying distribution $p(k)$, it has been shown that
an evolutionary growth procedure  with preferential attachment to
high degree nodes  leads to scale-free behavior \cite{prefattach}.
Furthermore, for many applications scale-free and small world networks 
emerge naturally upon optimization for
minimizing both the costs for communication and maintenance 
\cite{optimization} which are
central criteria for the design of PTNs.
 
Using scaling arguments for  tree graphs the dependence of a
number of properties of scale-free networks on the exponent
$\gamma$ has been worked out. In particular, the value of $\gamma$
discriminates between different classes for the percolation
behavior: if $\gamma>4$, the behavior is equivalent to that of
networks with exponential $p(k)$, while it is qualitatively
different for $\gamma<4$, moreover there is no percolation
threshold  for $2<\gamma<3$ \cite{Cohen02}. The percolation
threshold is given by: $\kappa\equiv\overline{k^2}/ \bar{ k}=2$. For the PTNs studied this ratio is listed in Table
\ref{tab1}. Finding a network with $\kappa$ near the threshold
means that it is vulnerable against failure in the sense that
inactivating a small number of nodes may break the network into
disconnected clusters. The Table shows that all PTNs are clearly
above this threshold, for some of them $\kappa$ indicates an
especially strong resilience against failure \cite{Ferber06b}.

%
%
\begin{figure}[ht]
\includegraphics[width=8cm]{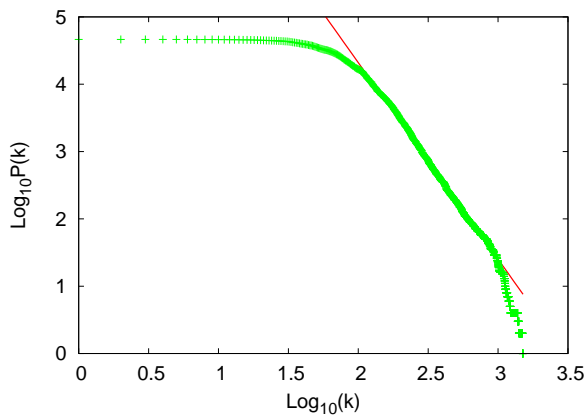}
   \caption{Integrated node degree distribution $P(k)=\int_k^{k_{\rm max}}p(k)dk$
   for the Los Angeles PTN in {\bf P}-space with fit to a power law.
   \label{fig2}}
\end{figure}

More information can be gained by fitting the degree distribution
to a power-law as above or to an exponential function $p(k)\sim
\exp(- k/\hat{k})$ observing the quality of the fit. In Table
\ref{tab1} we list the results of both fits for the {\bf L} and
{\bf P}-spaces. In the {\bf L}-space representation, the $p(k)$ of
eight cities is found to be well fitted by a power law with
$\gamma$ values between 2.6 and 5. The corresponding results for
the exponential fits for these cities are given in parentheses.
The other six cities are rather governed by an exponentially
decaying $p(k)$ (for these, the power law fit is shown in
parentheses). Whereas the scale-free behavior in the {\bf L}-space
has also been seen in previous work \cite{Sienkiewicz05,Ferber05},
such behavior was ruled out for the {\bf P}-space
\cite{Sienkiewicz05}. Fig. \ref{fig2} however clearly indicates
that the {\bf P}-space node degree distribution of the Los Angeles
PTN follows a power-law with exponent $\gamma_{\rm p}=3.9$. The
same also applies to London and Paris, all other cities appear to
have an exponentially decaying $p(k)$ in this space. Whereas an
exponential distribution may be explained by a random placement of
the routes, the power law found for the three cities indicates
that the routes are organized in a correlated way. The deeper
reasons for this special behavior currently remain unclear.

\begin{figure}[ht]
\parbox{7cm}{\includegraphics[width=7cm]{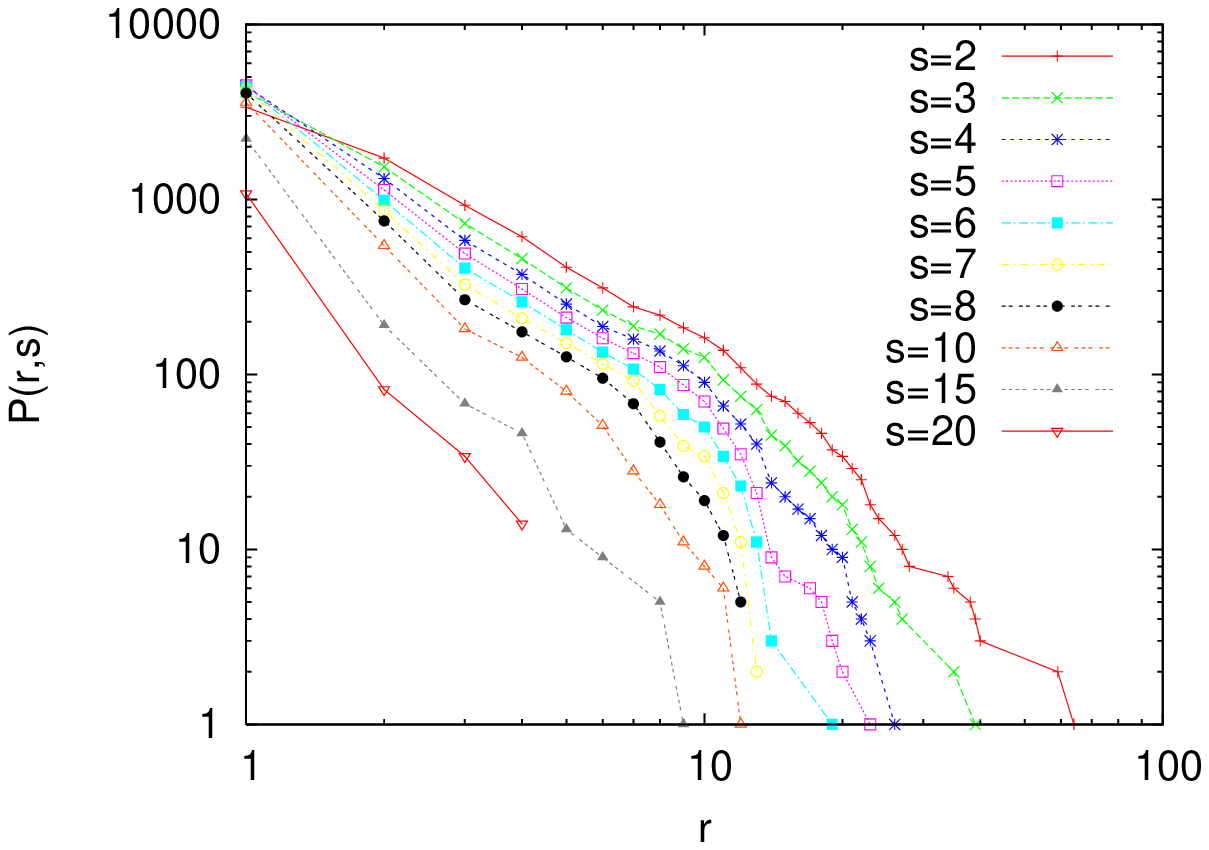}}{\bf (a)}\\
\parbox{7cm}{\includegraphics[width=7cm]{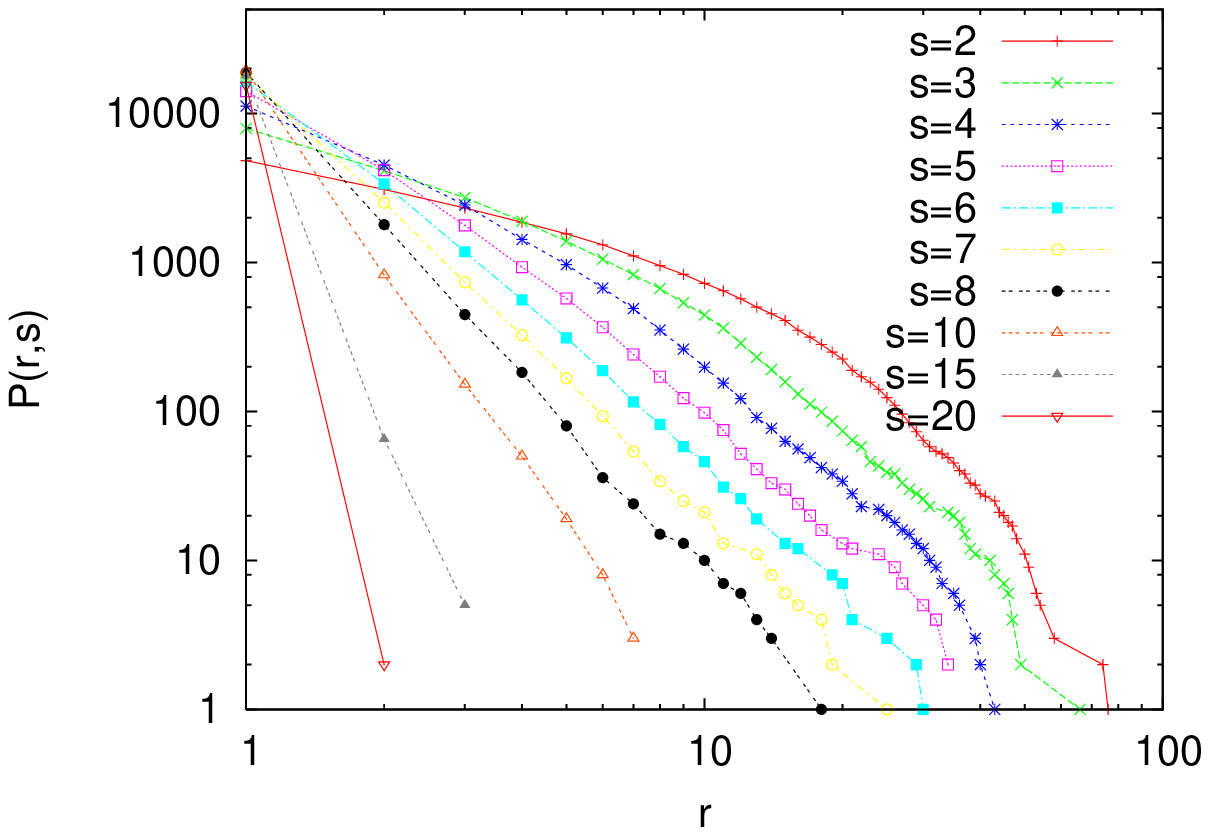}}{\bf (b)}
   \caption{Harness distribution: number of sequences of $s$ consecutive
stations that are serviced by $r$ parallel routes. a: Sydney PTN ,
b: simulated city with $a=0, b=0.5$.
   \label{fig3}}
\end{figure}

Using the full information about the route paths as included in
the {\bf L$^{'}$}-representation with colored links (Fig.
\ref{fig1}), we can extract specific PTN features. One of them is that several routes may proceed in
parallel on a common road or track for a sequence of stations. The
emerging picture very much resembles networks found in car wiring
technology, where the term {\em harness} is used. To
quantitatively describe this characteristic we introduce the harness
distribution $P(r,s)$: the number of sequences of $s$ consecutive
stations that are serviced by $r$ parallel routes. In Fig.
\ref{fig3}a we show the harness distribution for the PTN of Sydney. The
log-log plot indicates that also this distribution appears to be
scale-free. Similar behavior was found for most of the cities
included in our study. 
%
%

The evidence for the scale-free properties of PTNs presented so far
encourages us to propose an evolutionary growth model for these
networks along the following lines: We model the grid of streets
by a quadratic 2D lattice and allow every lattice site $\vec{r}$ (street
corner) to be a potential station visited by e.g. $k_{\vec{r}}$ routes. The routes are  modeled as
self-avoiding walks (SAWs) on this lattice. 
This model captures the typical meandering of the routes to optimize
the the coverage of neighborhoods. Obviously, real routes are also planned
to access sites of interest. These serve as points of attraction of the walks
and `integrating out' their specific locations one is left with an effective
attraction between the routes.  

\begin{figure}[ht]
\includegraphics[width=7cm]{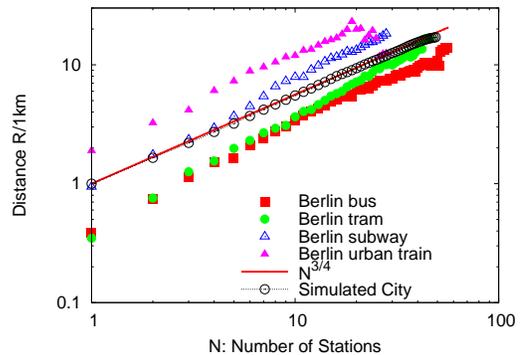}
   \caption{Berlin PTN: $R$ as function of the number of stations
traveled compared with the 2D SAW and a simulated city with
parameters $a=0, b=0.5$. \label{fig4}}
\end{figure}

The rules of our model are the
following:
\begin{itemize}
 \item [1.] First route: construct a SAW of length $n$.
 \item [2.] Subsequent routes: (i) choose a terminal station at the
 lattice site $\vec{r}$ with probability $q\sim k_{\vec{r}} +
 a$; (ii) choose a subsequent station of this route at a neighboring site
 site $\vec{r}'$ with probability $q\sim k_{\vec{r}'} + b$;
 (iii) repeat step (ii) until the walk has reached $n$  stations,
 in case of self-intersection discard the walk and restart with
 step (i).
 \item [3.] Repeat step 2 until $m$ routes are created.
\end{itemize}
Implementing the model on a 2D lattice implies the assumption of a
PTN growing in a regular isotropic uncorrelated environment. While
it is natural that the routes should not intersect themselves, the
hypothesis that apart from the effective attraction they otherwise proceed randomly may not be
obvious. SAWs are well studied in physics to model scaling
properties of  polymers and polymer networks \cite{saw1}. In 2D,
the end-to-end distance $R$  of a SAW of $N$ steps is known to
scale for large $N$ as $R\sim N^{\nu}$ with an exact result for
the exponent $\nu=3/4$ \cite{saw2}. This result as well as other
scaling properties remain unperturbed even on a weakly disordered
lattice as long as the disorder is not long-range correlated
\cite{saw3}. This supports our choice of disregarding such
possible disturbances of the lattice in our model. We have tested
this hypothesis using publicly available geographical data
\cite{note2} for the stations of the Berlin PTN. Plotting the root
mean square distance  $R$ as function of the number of stations
traveled starting from a terminal we find surprising agreement
with the 2D SAW behavior (see Fig. \ref{fig4}). Furthermore, the
corresponding result for a simulated city confirms this picture.
Apparently, the SAW balances between area coverage and traveling time.

%
%

In Table \ref{tab1} we have included  the characteristics as
extracted from simulated PTNs with $m=500$ routes each of $n=50$
stations for typical parameters $a$ and $b$. In particular our
model is nicely suited to reproduce the harnessing effect of the
PTNs as shown in Fig. \ref{fig3}b. Varying the parameter $b$ for
$a=0$ we observe a crossover from scale-free behavior in {\bf P}-space 
for small
$b$ to an exponential one as $b$ increases beyond 1. 
Some {\bf L}-space characteristics we left blank in the table due 
to artifacts of the square lattice neighborhood. 
Although our
growth rules look very similar to usual network evolution by
preferential attachment \cite{prefattach}, there is no simple
relation between the parameters $a$ and $b$ and the exponent
$\gamma$ even in the scale-free scenario. The principal difference
of our algorithm is that at each step we link an {\em existing}
station to a neighboring site which does not need to be empty. New
stations are then only added at the frontier of the PTN cluster
while high degree nodes (hubs) accumulate at its center.

Our analysis of PTNs of so far unexplored sizes brings about that very
large PTNs may display scale free {\bf P}-space distributions and
confirms corresponding {\bf L}-space results
\cite{Sienkiewicz05,Ferber05}.  The surprising SAW behavior of the
routes encouraged us to analyze an evolutionary  model of mutually 
attractive SAWs which reproduces a number of key features of PTNs.
In conclusion, we want to emphasize the
importance of the constraints that are imposed on the network by
the area consuming links when it is embedded in a 2D space. The
harnessing effect presented in this study being a first example.
Similar problems are met e.g. for electric circuit design.

We acknowledge support by the  EC under the Marie Curie Host
Fellowships for the Transfer of Knowledge, project COCOS, contract
MTKD-CT-2004-517186 (C.v.F.), and Austrian Fonds zur F\"orderung
der wi\-ssen\-schaft\-li\-chen Forschung under Project P16574
(Yu.H.).


\begin{thebibliography}{99}
\bibitem{reviews} R. Albert and A.-L. Barab\'asi,
Rev. Mod. Phys. {\bf 74}, 47 (2002); S.N. Dorogovtsev and J.F.F. 
Mendes, Adv. Phys. {\bf 51}, 1079 (2002);  M.E.J. Newman,
SIAM Review {\bf 45}, 167 (2003); S.N. Dorogovtsev and J.F.F. 
Mendes, {\em Evolution of Networks} (Oxford University Press,
Oxford, 2003).
\bibitem{Amaral00} L.A.N. Amaral, A. Scala, M. Barth\'el\'emy,
and H.E. Stanley,  Proc. Nat. Acad. Sci. USA., {\bf 97}, 11149
(2000).
\bibitem{air} R. Guimer\`a and L.A.N. Amaral, Eur. Phys. J. B {\bf 38},
381 (2004);  R. Guimera, S. Mossa, A. Turtschi, and L.A.N. Amaral,
Proc. Nat. Acad. Sci. USA {\bf 102}, 7794 (2005);  A. Barrat, M.
Barth\'elemy, R. Pastor-Satorras, and A. Vespignani, Proc. Nat.
Acad. Sci. USA {\bf 101}, 3747 (2004); W. Li and X. Cai, Phys.
Rev. E {\bf 69}, 046106 (2004); W. Li, Q.A. Wang, L. Nivanen, and
A. Le M\'ehaut\'e,  e-print physics/0601091.
\bibitem{Sen03} P. Sen, S. Dasgupta, A. Chatterjee, P.A. Sreeram,
G. Mukherjee, and S.S. Manna, Phys. Rev. E {\bf 67}, 036106
(2003)
\bibitem{powergrid} P. Crucitti, V. Latora, and M. Marchiori,
Physica A {\bf 338}, 92 (2004); R. Albert, I. Albert, and G.L.
Nakarado, Phys. Rev. E {\bf 69}, 025103(R) (2004).
\bibitem{Marchiori00} M. Marchiori and V. Latora, Physica A {\bf 285}, 539 (2000).
\bibitem{Latora} V. Latora and M. Marchiori, Phys. Rev. Lett. {\bf 87},
198701 (2001); V. Latora and M. Marchiori, Physica A {\bf 314},
109 (2002).
\bibitem{Seaton04} K.A. Seaton and L.M. Hackett, Physica A {\bf 339},
635 (2004).
\bibitem{Sienkiewicz05} J. Sienkiewicz and J.A. Ho\l{}yst, Phys. Rev. E {\bf 72}, 046127
(2005).
\bibitem{Ferber05} C. von Ferber, Yu. Holovatch, and V. Palchykov,
Condens. Matter Phys. {\bf 8}, 225 (2005).
\bibitem{Ferber06b} C. von Ferber, T. Holovatch, Yu. Holovatch, and V. Palchykov,
in preparation.
\bibitem{note1} For links see: http://www.apta.com.
\bibitem{Watts98} D.J. Watts, S.H. Strogatz, Nature {\bf 393}, 440
(1998).
\bibitem{prefattach} A.-L. Barab\'asi and R. Albert,
Science {\bf 286}, 509 (1999);  A.-L. Barab\'asi, R. Albert, and
H. Jeong, Physica A {\bf 272}, 173 (1999).
\bibitem{optimization}  
S. Valverde, R. Ferrer Cancho, and R.V. Sol\'e, Europhys. Lett. {\bf 60}, 512 (2002);  
M.T. Gastner and M.E.J. Newman, Eur. Phys. J. B {\bf 49}, 247 (2006);
N. Mathias and V. Gopal, Phys. Rev. E {\bf 63}, 021117 (2001).
\bibitem{Cohen02} R. Cohen, D. ben-Avraham, and S. Havlin,
Phys. Rev. E {\bf 66}, 036113 (2002).
\bibitem{saw1}
P.-G.~de~Gennes, {\em Scaling Concepts in Polymer Physics}
(Cornell University Press, Ithaca and London, 1979).
\bibitem{saw2}
B. Nienhuis, Phys. Rev. Lett. {\bf 49}, 1062 (1982).
\bibitem{saw3} A.B. Harris,  Z. Phys. B  {\bf 49}, 347 (1983);
V. Blavats'ka, C. von Ferber, and Yu. Holovatch, Phys. Rev. E,
{\bf  64},  041102  (2001); C. von Ferber and Yu. Holovatch,
Phys. Rev. E {\bf 65}, 042801 (2002); C. von Ferber, V.
Blavats'ka, R. Folk, and Yu. Holovatch, Phys. Rev. E {\bf  70},
035104(R) (2004).
\bibitem{note2} For maps see: http://www.fahrinfo-berlin.de.
 \end{thebibliography}
\end{document}